\documentstyle[11pt,amssymb,epsf]{article}

\def\dalemb#1#2{{\vbox{\hrule height .#2pt
        \hbox{\vrule width.#2pt height#1pt \kern#1pt
                \vrule width.#2pt}
        \hrule height.#2pt}}}

\def\0{{\sst{(0)}}}
\def\1{{\sst{(1)}}}
\def\2{{\sst{(2)}}}
\def\3{{\sst{(3)}}}
\def\4{{\sst{(4)}}}
\def\5{{\sst{(5)}}}
\def\6{{\sst{(6)}}}
\def\7{{\sst{(7)}}}
\def\8{{\sst{(8)}}}

\def\wtd{\widetilde}

\def\nn{\nonumber} \def\bd{\begin{document}} \def\ed{\end{document}}
\def\ds{\documentstyle} \let\fr=\frac \let\bl=\bigl \let\br=\bigr
\let\Br=\Bigr \let\Bl=\Bigl 
\let\bm=\bibitem
\let\na=\nabla
\let\pa=\partial \let\ov=\overline 
\newcommand{\be}{\begin{equation}} 
\newcommand{\ee}{\end{equation}} 
\def\ba{\begin{array}}
\def\ea{\end{array}}
\def\ft#1#2{{\textstyle{{\scriptstyle #1}\over {\scriptstyle #2}}}}
\def\fft#1#2{{#1 \over #2}}
\def\del{\partial}
\def\sst#1{{\scriptscriptstyle #1}}
\def\oneone{\rlap 1\mkern4mu{\rm l}}
\def\ie{{\it i.e.\ }}
\def\etc{{\it etc.\ }}
\def\via{{\it via}}
\def\semi{{\ltimes}}
\def\cv{{\cal V}}
\def\str{{\rm str}}
\def\jm{{\rm j}}
\def\im{{\rm i}}

\def\cramp{\medmuskip = 2mu plus 1mu minus 2mu}
\def\cramper{\medmuskip = 2mu plus 1mu minus 2mu}
\def\crampest{\medmuskip = 1mu plus 1mu minus 1mu}
\def\uncramp{\medmuskip = 4mu plus 2mu minus 4mu}

\def\mapright#1{\smash{\mathop{-\!\!\!-\!\!\!-\!\!\!-\!\!\!-\!\!\!
             \longrightarrow}\limits^{#1}}}
\def\maprightt#1#2{\smash{\mathop{-\!\!\!-\!\!\!-\!\!\!-\!\!\!-\!\!\!
             \longrightarrow}\limits^{#1}_{#2}}}

\def\tX{{{\wtd X}}}
\newcommand{\ho}[1]{$\, ^{#1}$}
\newcommand{\hoch}[1]{$\, ^{#1}$}
\newcommand{\bea}{\begin{eqnarray}} 
\newcommand{\eea}{\end{eqnarray}} 
\newcommand{\ra}{\rightarrow}
\newcommand{\lra}{\longrightarrow}
\newcommand{\Lra}{\Leftrightarrow}
\newcommand{\ap}{\alpha^\prime}
\newcommand{\bp}{\tilde \beta^\prime}
\newcommand{\tr}{{\rm tr} }
\newcommand{\Tr}{{\rm Tr} } 
\newcommand{\NP}{Nucl. Phys. }
\newcommand{\tamphys}{\it Center for Theoretical Physics\\
Texas A\&M University, College Station, Texas 77843}
\newcommand{\upenn}{\it Department of Physics and Astronomy\\
University of Pennsylvania, Philadelphia, Pennsylvania 19104}

\newcommand{\auth}{M. Cveti\v{c}\hoch{\dagger1}, 
H. L\"u\hoch{\dagger1} and C.N. Pope\hoch{\ddagger2}
 }

\begin{document}
\begin{flushright}
\hfill{CTP TAMU-01/00}\\
\hfill{UPR-868-T}\\
\hfill{hep-th/0001002}\\
\hfill{December 1999}\\
\end{flushright}

\vspace{15pt}

\begin{center}
{ \large {\bf Domain Walls and Massive Gauged Supergravity Potentials
}}

\vspace{15pt}
\auth

\vspace{15pt}

{\hoch{\dagger}\upenn}

\vspace{15pt}
{\hoch{\ddagger}\tamphys}

\vspace{40pt}

\underline{ABSTRACT}
\end{center}

We point out that massive gauged supergravity potentials, for example
those arising due to the massive breathing mode of sphere reductions
in M-theory or string theory, allow for supersymmetric (static) domain
wall solutions which are a hybrid of a Randall-Sundrum domain wall on
one side, and a dilatonic domain wall with a run-away dilaton on the
other side.  On the anti-de Sitter (AdS) side, these walls have a
repulsive gravity with an asymptotic region corresponding to the
Cauchy horizon, while on the other side the runaway dilaton approaches
the weak coupling regime and a non-singular attractive gravity, with
the asymptotic region corresponding to the boundary of space-time.  We
contrast these results with the situation for gauged supergravity
potentials for massless scalar modes, whose supersymmetric AdS extrema
are generically maxima, and there the asymptotic regime transverse to
the wall corresponds to the boundary of the AdS space-time. We also
comment on the possibility that the massive breathing mode may, in the
case of fundamental domain-wall sources, stabilize such walls via a
Goldberger-Wise mechanism.

{\vfill\leftline{}\vfill
\vskip 5pt
\footnoterule
{\footnotesize \hoch{1} Research supported in part by DOE grant 
DE-FG02-95ER40893 \vskip -12pt} \vskip 14pt
%{\footnotesize \hoch{2} Research supported in part by DOE grant
%DE-AC02-76ER03071 \vskip -12pt} \vskip 14pt
{\footnotesize  \hoch{2} Research supported in part by DOE 
grant DE-FG03-95ER40917.\vskip  -12pt}}

\pagebreak
\setcounter{page}{1}

The scalar potentials of gauged supergravity theories provide a
natural testing-ground for studying domain-wall configurations within
the framework of a basic theory.  In general, such scalar potentials
have isolated supersymmetric extrema with a negative cosmological
constant.  Within the AdS/CFT correspondence, these supersymmetric
(BPS) domain walls play a role in elucidating the renormalization
group flows and bound-state spectra of strongly coupled gauge theories
(see, for example, \cite{010}-\cite{deBVV} and references therein).  A
typical feature of gauged supergravity potentials is such that the
supersymmetric extrema are {\it maxima} of the potential.  The domain
walls are therefore typically those with {\it negative tension}, and
the metric transverse to the wall asymptotically ($z\to \infty$)
approaches the {\it boundary of the AdS space-time} \cite{BC}. Another
feature of these solutions is that the region near the wall ($z\to 0$)
is in general singular; both the scalar field and the curvature
generically exhibit singular behavior and thus the continuation across
the wall region on the side $z<0$ involves (within the effective
theory) a continuation across a singular domain-wall regime
(c.f. \cite{BC,BCp}).

On the other hand, in recent months there has been a resurgence in the
study of domain walls in asymptotically AdS space-times in $D=5$
gravity theories.  For special examples of such static domain walls,
the gravity effects transverse to the wall are suppressed, which has
interesting implications for the phenomenology of the world on the
brane.  (See, for example, \cite{050}-\cite{260} and references
therein.)  Non-static walls in $D=5$ were also recently considered.
(See \cite{170}-\cite{CW} and references therein.)~\footnote{It turns
out ~\cite{CW} that the local and global space-time structure of
vacuum domain walls ($(D-2)$-brane configurations) in $D$ dimensions
is universal, and thus the previous studies of domain walls in $D=4$
(see, \cite{CGS,CS} and references therein) are completely parallel to
the domain-wall solutions in any other dimension $D$.}

A particular focus is on infinitely thin, static, $Z_2$-symmetric
domain-wall solutions, constructed \cite{050,070} in a pure AdS
gravity theory (the Randall-Sundrum scenario).\footnote{Another
proposal for the origin of the five-dimensional domain wall was made
in \cite{losw}, which is dilatonic and can be viewed as M5-branes
wrapped arond the two-cycles of a Calabi-Yau manifold.}
(Generalizations that incorporate the effects of additional
compactified dimensions were given in \cite{060,250,260}.) These
solutions have a repulsive gravity \cite{CG}, for which the asymptotic
regions ($z\to \pm \infty$) corresponding Cauchy horizons
\cite{CDGS,Gibb}. They satisfy \cite{050,070} a specific relation
between the domain-wall tension $\sigma$ and the cosmological constant
$\Lambda$ of the AdS vacuum; this latter condition was subsequently
shown \cite{BC} to be a consequence of supersymmetry.  (These results
are again completely parallel \cite{CGR} with supersymmetric domain
walls of N=1 supergravity theories in $D=4$.)  These types of wall are
of Type II in the classification scheme of refs.~\cite{CG,BC}.

The main motivation of this paper is to provide a framework within
gauged supergravity theories that has a chance of implementing the
Randall-Sundrum scenario(s).  As mentioned above, gauged supergravity
theories tend to have potentials for the massless scalar modes that
have isolated supersymmetric maxima and {\it not} minima.  Thus the
supersymmetric domain walls have negative tension (whose magnitude is
the same as the tension of Type II walls).  They have attractive
gravity transverse to the wall, with the asymptotic regions ($z\to \pm
\infty$) corresponding to AdS space-time boundaries
\cite{BC,BCp}. These types of walls are referred to as Type IV walls
\cite{BC} and are complementary to Type II walls.
 
In order to obtain Type II domain-wall solutions of the
Randall-Sundrum scenario, the gauged supergravity potential would have
to have two isolated supersymmetric {\it minima}. Since the potentials
for the massless scalar fields in a gauged supergravity do not seem to
have this feature, we turn in our analysis to include other scalar
fields that do not lie in the massless supermultiplet.

We shall focus on the special classes of gauged supergravities that
arise from sphere reductions of M-theory or string theory, with
particular emphasis on the $D=5$ case.  For examples in the
Kaluza-Klein reduction of Type IIB string theory on a five-sphere
($S^5$), there will be an infinite tower of massive supermultiplets in
addition to the massless multiplet, and so one could consider the
potentials for one or more of the massive scalar fields.  In general,
one cannot focus attention on a single such field in isolation, on
account of its couplings to other fields.  However, in certain special
cases a consistent truncation to a single massive scalar can be
performed.  One such example is the ``breathing mode'' that
parameterises the overall volume of the compactifying $S^5$.  (Unlike
the breathing mode in a toroidal reduction, which is massless, the
breathing mode in a spherical reduction is a member of a massive
supermultiplet.)

   The scalar potentials for the breathing-mode scalars in various
Kaluza-Klein spherical reductions were studied in \cite{bdlps}.
Although the breathing mode is a member of a massive multiplet, the
truncation is nonetheless consistent since it is a singlet under the
isometry group of the internal sphere. (It would not in general be
consistent to turn on a finite subset of other fields as well.)

The resulting $D$-dimensional
Lagrangians all turn out to have the following form:
%%%%%
\be
{\cal L}_D = e\, R - \ft12 e\, (\del\phi)^2 - e\, V\,,
\ee
%%%%%
where the potential is given by \cite{bdlps}
%%%%%%%%
\be
V=\ft12\, g^2(\fft{1}{a_1^2}\, e^{a_1\phi} - \fft{1}{a_1 a_2}\, 
e^{a_2\phi})\,.\label{scalarpot}
\ee
%%%%%%%
The positive constants $a_1$ and $a_2$ are given by
%%%%%
\be
a_1^2 = \fft{4}{N} + \fft{2(D-1)}{D-2}\,,\qquad a_1\, a_2=
\fft{2(D-1)}{D-2}\,,
\ee
%%%%%%
where $N$ is a certain positive integer.  For $D=4$, 7 and 5, this
integer takes the value $N=1$.  These cases correspond to the $S^7$
and $S^4$ reductions of $D=11$ supergravity, and the $S^5$ reduction
of type IIB supergravity respectively.  For $D=3$ the integer $N$ can
be equal to 1, 2 or 3, corresponding to the $S^1$ reduction of the
Freedman-Schwarz model,\footnote{The reduction in this case is of a
generalised Scherk-Schwarz type, where the axion is allowed a linear
dependence on the reduction coordinate.} the $S^3$ reduction of $D=6$
simple (chiral) supergravity, and the $S^2$ reduction of $D=5$ simple
supergravity respectively.   The explicit dilaton coupling constants
$a_1$ and $a_2$ for the above cases are given in Table 1.
%%%%%%%%%
\begin{center}
\begin{tabular}{|c|c|c|c|c|}
\hline D& N&$a_1$& $a_2$&${a_1\over {a_2(D-1)}}$\\\hline\
7& 1& $4\sqrt{2\over 5}$ & $3\over \sqrt{10}$   &${4\over 9}$\\
5& 1& $2\sqrt{5\over 3}$ & ${4\over \sqrt{15}}$ &${5\over 8}$\\
4& 1& $\sqrt7$           & ${3\over \sqrt{7}}$  &${7\over 9}$\\
3& 1& $2\sqrt2$          & $\sqrt2$             &$1$ \\
3& 2& $\sqrt6$           & $2\sqrt{2\over3}$    &${3\over 4}$\\
3& 3& $4\over \sqrt3$    &$\sqrt3$              &${2\over 3}$\\
\hline
\end{tabular}
\end{center}
\bigskip

\centerline{Table 1: The values of the parameters 
$N$, $a_1$ and $a_2$ in diverse dimensions $D$}
\noindent{\phantom{xxxxxxxxxxxxx}
that enter the scalar potential (\ref{scalarpot}).}

\bigskip

 Since  $a_1>a_2>0$, the
potential has a minimum at $\phi=0$, with
%%%%%
\be
V_{\rm min} = -\fft{g^2\, (D-2)}{N\, (D-1)\, a_1^2}\,.
\ee
%%%%%
See Figure 1 for the shape of the potential.  The potential can be
expressed in terms of a ``superpotential'' $W_+$ or $W_-$ as follows:
%%%%
\be
V=(\fft{\del W_\pm}{\del \phi})^2 -\fft{D-1}{2(D-2)}\, W_\pm^2\,,
\ee
%%%%%
where
%%%%
\be
W_\pm=\sqrt{\fft{N}{2}}\, g\, (\fft{1}{a_1}\, e^{a_1\phi/2} \pm 
\fft1{a_2}\, e^{a_2\phi/2})\,.
\ee
%%%%%

\begin{figure}[ht]
\leavevmode\centering
\epsfbox{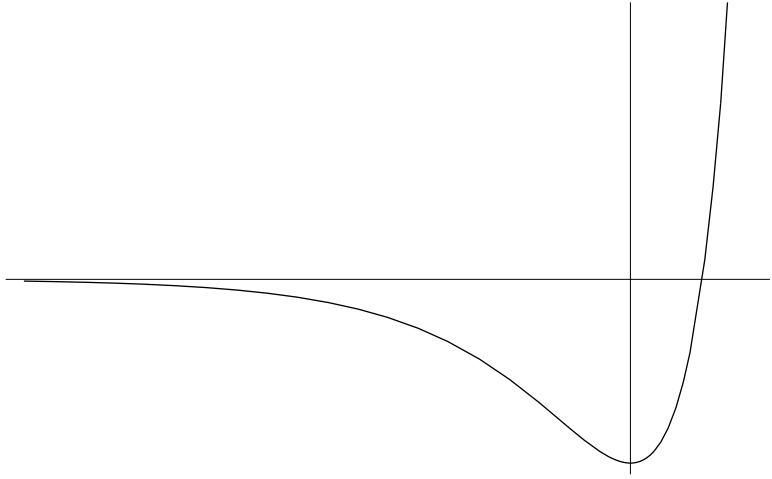}
\caption{The scalar potential (\ref{scalarpot}) as a function of
$\phi$ for  the $D=5$ case. The structure of the potential for other 
dimensions is similar.}
\end{figure}

       Let us now consider the following ansatz for a domain-wall
metric:
%%%%%
\be
ds^2 = e^{2A}\, dx^\mu\, dx^\nu\, \eta_{\mu\nu} + dz^2\,.
\ee
%%%%
The equations of motion are
%%%%
\bea
&&\phi'' + (D-1)\, A'\, \phi'= \fft{\del V}{\del \phi}\nn\\
&&A'' + (D-1) (A')^2=(D-1)\, A'' + (D-1)\, (A')^2 +
\ft12 (\phi')^2 = - \fft{V}{D-2}\,.\label{eom}
\eea
%%%%%%%
These admit a first integral, given by
%%%%%%
\be
\phi' = \sqrt2 \fft{\del W_\pm}{\del \phi}\,,\qquad 
A' = -\fft{1}{\sqrt2\, (D-2)}\, W_\pm\,.\label{firstint}
\ee
%%%%%

  Here, we shall consider the choice $W_-$ for the superpotential,
since it has a supersymmetric minimum, i.e. $\partial_\phi W_-=0$, at
$\phi=0$.  From (\ref{firstint}), we shall therefore have
%%%%%
\bea
\phi'&=&\sqrt{\fft{N}{4}}\, g\, (e^{a_1\phi/2} -e^{a_2\phi/2})\,,\nn\\
A'&=& =-  \fft{g\, \sqrt N}{2(D-2)}\, 
[\fft{1}{a_1} e^{a_1\phi/2} -\fft{1}{a_2} e^{a_2\phi/2}]\,.
\eea
%%%%%

    Solving for $\phi$ and $A$, we find that $A$ can be expressed as a
function of $\phi$, namely
%%%%%
\be
e^{(D-1)\, A} =  c\, \fft{\del W}{\del \phi}\, e^{-\fft12
(a_1+a_2)\phi}\,,\label{afromphi}
\ee
%%%%%
where $c$ is an integration constant.  For $D>3$, the solution for
$\phi$ is given by
%%%%%
\be
z-z_0 = \fft4{a_2\, g\, \sqrt N}\, e^{-\fft12 a_2\, \phi}\, 
{}_2F_1[\fft{a_2}{a_2-a_1}, 1, 1+\fft{a_2}{a_2-a_1}; e^{\fft12
(a_1-a_2)\, \phi}]\,.\label{hype}
\ee
%%%%%
(For our specific examples
mentioned above, we shall have $N=1$ and $D=4$, 5 or 7.)  For $D=3$,
we find that $\phi$ is given by
%%%%%
\bea
N=1:&& z-z_0= \fft{\sqrt8}{g}\, \Big( e^{-\fft1{\sqrt2}\, \phi} +
\log( e^{-\fft1{\sqrt2}\, \phi}-1)\Big)\,,\nn\\
N=2:&& z-z_0= \fft{\sqrt3}{g}\, \Big( e^{-\sqrt{\fft23}\, \phi} +
+ 2 e^{-\fft1{\sqrt6}\, \phi} + 2 
\log( e^{-\fft1{\sqrt6}\, \phi}-1)\Big)\,,\label{d3sol}\\
N=3:&& z-z_0= \fft{2}{g}\, \Big( e^{-\fft1{\sqrt3}\, \phi} +
\ft23 e^{-\fft{\sqrt3}{2}\, \phi}
+ 2 e^{-\fft1{2\sqrt3}\, \phi} + 2 
\log( e^{-\fft1{2\sqrt3}\, \phi}-1)\Big)\,,\nn
\eea
%%%%%
(The analogous solutions constructed using $W_+$ rather than $W_-$ can
also be easily obtained, but they seem not to be directly relevant for
our present purposes.)  These supersymmetric domain walls in a
different coordinate system were given in \cite{bdlps}, where their
higher dimensional origins as M-branes and D3-branes were discussed.
Note that the hypergeometric function ${}_2F_1[a, 1, 1+a, x]$ appearing
in (\ref{hype}) is the
Lerch transcendent $a\,\Phi(x, 1, a)$.  In fact, the solutions for
$D>3$ and for $D=3$ can all be given by a single formula using the
Lerch transcendent, namely
%%%%%
\be
z-z_0 = -\fft{a_1\, \sqrt{N}}{g}\, e^{-\fft12 a_2\, \phi}\, 
\Phi(e^{\fft12 (a_1-a_2)\, \phi}, 1, \fft{a_2}{a_2-a_1})\,.
\ee
%%%%%

    The $W_-$ solutions above all have two different branches. In one
branch, $\phi$ runs from 0 to $+\infty$, with $z$ runnning from
$z=-\infty$ to $z=0$, where we have chosen the integration
constant $z_0$ to be
%%%%%
\bea
D>3:&&z_0 = \fft{\pi\, a_1\, \sqrt{N}}{g}\, \Big( -\im + \cot(\fft{\pi\,
a_1}{a_1-a_2})\Big)\,,\nn\\
D=3:&& z_0 = \fft{\pi\, a_1\, \sqrt{N}}{g}\,.
\eea
%%%%%
(The imaginary part cancels the imaginary
additive constant in the expressions (\ref{hype}) and (\ref{d3sol}).)   
When $\phi$ is large, the solution takes the form
%%%%%
\bea
e^{-\fft12 a_1\, \phi} &\sim& -\ft14 a_1\,\sqrt{N}\, g\, z\,,\nn\\
e^{(D-1)\, A} &\sim& c\, \sqrt{\fft{N}{8}}\, g\, e^{-\fft12 a_2\,
\phi} \sim  c\, \sqrt{\fft{N}{8}}\, g\,\Big(- \ft14 a_1\, \sqrt{N}\,
g\, z\Big)^{\fft{a_2}{a_1}}\,.
\eea
%%%%%
In this branch, when the coordinate $z$ reaches its limit at $z=0$,
the factor $e^{2A}$ in the metric therefore goes to zero, and there is a
power-law naked curvature singularity.   
(Note that in this regime the solution  extends
into  large positive values of the  potential (\ref{scalarpot}) with a  
large cost to the energy density of the wall, and it
thus terminates at a finite value of the transverse coordinate.)

As $z$ approaches $-\infty$, the
functions $\phi$ and $A$ become
%%%%%
\be
\phi \sim e^{\fft{g}{a_1\, \sqrt{N}}\, z}\,,\qquad 
A\sim \fft{g}{a_1\, \sqrt{N}\, (D-1)}\, z\,.
\ee
%%%%%
The metric  asymptotically approaches the  AdS  space-time, described in
horospherical coordinates with $z\to -\infty$ corresponding to the Cauchy
horizon \cite{CDGS,Gibb}. Note that on that side of the wall the gravity is
repulsive and  provides ``one half'' of the Randall-Sundrum wall.

    In the second branch, $\phi$ runs from 0 to $-\infty$, while $z$
runs from $z=-\infty$ to $z=+\infty$.  The behaviour of the solution
near $z=-\infty$ is the same as in the branch discussed previously,
with the metric approaching  asymptotically AdS. 
 As $z$ approaches $+\infty$, the
solution becomes
%%%%%
\bea
e^{-\fft12 a_2\, \phi} &\sim& \ft14 a_2\,\sqrt{N}\, g\, z\,,\nn\\
e^{(D-1)\, A} &\sim& - c\, \sqrt{\fft{N}{8}}\, g\, e^{-\fft12 a_1\,
\phi} \sim  - c\, \sqrt{\fft{N}{8}}\, g\,\Big( \ft14 a_2\, \sqrt{N}\,
g\, z\Big)^{\fft{a_1}{a_2}}\,.
\eea
%%%%%
(The constant $c$ is negative in this case.)  This side describes
one-side of a supersymmetric {\it dilatonic} domain wall \cite{C}.
Interestingly, it has {\it no curvature singularity}; as $z$ tends to
$+\infty$ the curvature falls off as $1/z^2$, while the diverging
dilaton $\phi\to -\infty$ approaches the weak coupling limit.  Gravity
on this side is attactive and for the null geodesics the affine
parameter $\tau$ is {\it infinite}.  Namely, $\tau \sim
\int^{+\infty}\, e^{-A}\, dz\sim z^{1-{a_1\over
{a_2(D-1)}}}|^{+\infty}$. Since for all the cases under consideration
the ratio ${a_1\over {a_2(D-1)}}\le 1$ (see Table 1),
$\tau$ is indeed infinite and $z\to +\infty$ corresponds to the
boundary of the space-time. ($D=3$, $N=1$ case is borderline with the
affine parameter diverging logarithmically.)

   For the purpose of constructing a domain-wall universe, it is the
second of the two branches that is relevant.  Thus this solution is a
hybrid of the Type II vacuum domain wall and the dilatonic wall.  The
thickness of the wall is of the order of $1/g$. It is a non-singular
solution, with repulsive gravity on the AdS-side ($z<0$) and
attractive gravity one on the dilatonic-side.  While $z=-\infty$
corresponds to the AdS Cauchy horizons, $z\to +\infty$ is the
time-like boundary of the space-time.  The solutions for both the
metric coefficient $e^{2A}$ and the breathing-mode scalar $\phi$, as
functions of $z$, are sketched in Figure 2 below.

\bigskip

\begin{figure}[ht]
\leavevmode\centering
\epsfbox{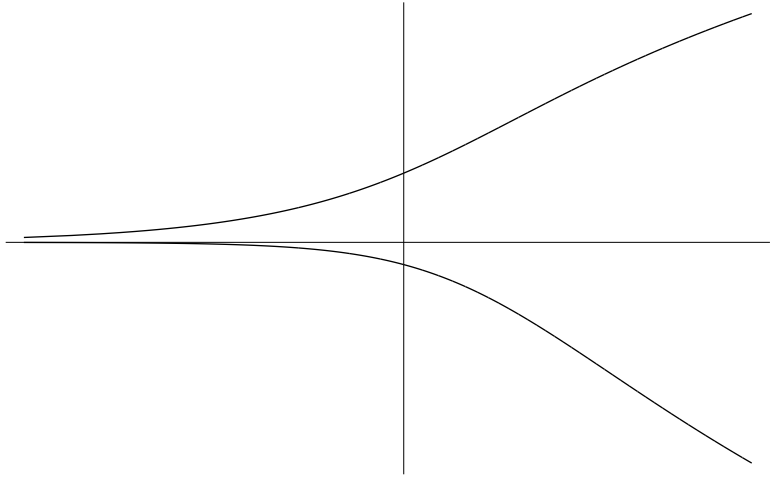}
\caption{The functions $e^{2A}$ (upper line) and $\phi$ (lower
line) for $D=5$ as functions of the transverse coordinate $z$. Their forms
for other dimensions are similar.}
\end{figure}     

Thus within a pure field-theoretic framework, \ie employing only the
breathing-mode scalar field to construct the domain wall solution, we
have been only partially successful; the massive gauged supergravity
potential gave us one ``supersymmetric AdS minimum'' and another
``run-away vacuum'', thus yielding a hybrid domain wall solution, and
not the pure Type II vacuum domain wall that we were really after.
Somewhat disappointing is the fact that on the dilatonic domain wall side
gravity is attractive, and thus these domain walls cannot provide a
phenomenologically viable scenario with a large transverse direction
$z=\{-\infty,+\infty\}$; only  the domain $z<0$ can be taken large. 

   We may also explore another possibility, by adding a singular
domain-wall source to this potential.  The breathing-mode potential
then provides a framework for implementing the Goldberger-Wise
scenario \cite{GW}.  In this case, in the second branch the diverging 
behaviour of
the dilaton is cancelled by a delta-function source for the domain
wall at some finite value of $z$, say $z= z_*$.  (Note that the source
tension has to precisely balance that of the scalar contribution at
the wall \cite{dWGFK}.)  Then, the solution for $z > z_*$ can be
replaced by a reflection of the solution for $z< z_*$, so that
%%%%%
\be
-|z -z_*| + z_* = \fft4{a_2\, g\, \sqrt N}\, e^{-\fft12 a_2\, \phi}\, 
{}_2F_1[\fft{a_2}{a_2-a_1}, 1, 1+\fft{a_2}{a_2-a_1}; e^{\fft12
(a_1-a_2)\, \phi}]\,.\label{hype2}
\ee
%%%%%
The metric function $A$ is again given by substituting $\phi$ into
(\ref{afromphi}).  (The reason why a solution can be constructed in
this way is because the original equations of motion (\ref{eom}) are
invariant under $z\longrightarrow -z$, and $z\longrightarrow
z+\,$constant.)  Since $A$ is continuous at $z=z_*$, but its first
derivative is not, it follows that there will be a delta-function
curvature singularity there.  This can be balanced by a domain-wall
source term, in precisely the same way that one can balance the
delta-function singularity on an electric string or $p$-brane soliton
with an appropriate source term.  The functions $e^{2A}$ and $\phi$ as
a function of $z$ for this solution are plotted in Figure 3 below.

\bigskip

\begin{figure}[ht]
\leavevmode\centering
\epsfbox{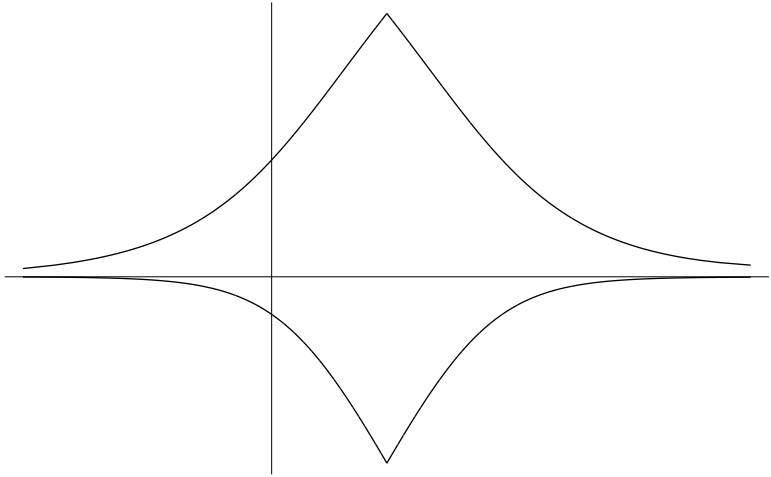}
\caption{The metric functions $e^{2A}$ (upper) and $\phi$ (lower) as a
function of the transverse direction $z$ with the domain wall source
added by hand. The solution provides a realization of the
Goldberger-Wise \cite{GW} mechanism, where the massive breathing mode
provides a potential that stabilizes the domain wall source and
relaxes asymptotically to the AdS {\it minimum} of the potential.}
\end{figure}

     To summarise, in this paper we set out to explore the possibility
of finding a supersymmetric AdS domain-wall solution, relevant for
$D=5$ for the Randall-Sundrum scenario, within massive gauged
supergravity theories.   By employing the potential for the massive
breathing-mode scalar of the compactifying sphere in M-theory or
string theory in diverse dimensions, we arrived at static
(supersymmetric) domain walls which are of a hybrid type.  On one
side they correspond to the Randall-Sundrum wall with repulsive
gravity, and on the other side they are supersymmetric dilatonic walls
\cite{C}.

   Although these supergravity solutions {\it per se} do not possess
all of the features needed for a Randall-Sundrum scheme,
one can obtain a more satisfactory result by including  also a 
fundmental (singular) domain-wall source.   The massive scalar mode acts as
a modulus stabilizing the domain wall (as in the Goldberger-Wise
scenario), and it provides a repulsive (AdS) gravity transverse to
the wall, as required for implementing the Randall-Sundrum
scenario.

\end{document}